# Magnetic Lens Made of a Single Solenoid for Controlling Bending of Two-Dimensional Ion Beam

Ardi Khalifah[1], Riri Murniati[2], Mikrajuddin Abdullah[3,4]

[1]SMA PGII 1, Bandung, Jl. Panatayuda 2, Bandung 40132, Indonesia

[2]Department of Military Physics, Indonesia Defense University, IPSC Sentul, Kabupaten Bogor 16810, Indonesia

[3]Department of Physics, Bandung Institute of Technology, Jalan Ganesa 10, Bandung, 40132, Indonesia

[4]E-mail: mikrajuddin@gmail.com

**Abstract**

The magnetic field inside an ideal solenoid cavity with an arbitrary cross-section is always constant, while it is always zero outside the solenoid. We can make a solenoid lens that can focus a parallel beam to a point behind it by adjusting the curvature of the solenoid circumference. In this paper, we discuss the design of magnetic lenses ranging from simple geometries to the general ones. We discovered that there are an infinite number of curvatures that can be used to focus the parallel beam to a specific focal point. Using this property, we also present the concept of a simple mass spectrometer by measuring the intensity of the ion captured by a detector placed at the focal point. This result is expected to enrich learning material in undergraduate courses, especially for the topic of electricity and magnetism.

## Introduction

Magnetic lenses have been used in a variety of devices, including cathode ray tubes, electron microscopes [1-3], ion beam lithography [4,5], accelerator [5], electron and ion beam emitted focusing systems [6] and others [7]. Magnetic lens designs are typically quite complex [8-10], necessitating the placement of electromagnets in quadrupole, sextupole, and higher pole structures.

This is done because the field's direction must be such that the charge beam entering it is bent to a specific direction.

In this paper, we will look at a simple design of a magnetic lens for focusing a two-dimensional particle beam, where the magnetic field is generated by an ideal solenoid. An ideal solenoid of any cross-section generates a constant magnetic field

$$B = \mu_0 n I \tag{1}$$

inside the solenoid cavity and of zero outside the solenoid [11]. Here $\mu_0$ represents the magnetic permeability in the solenoid cavity, $n$ represents the number of turns per unit length, and $I$ represents the current flowing through the solenoid. We can use this behavior to localize a space with a constant magnetic field inside a closed loop and a zero field outside the loop. The trick is to change the solenoid's cross section.

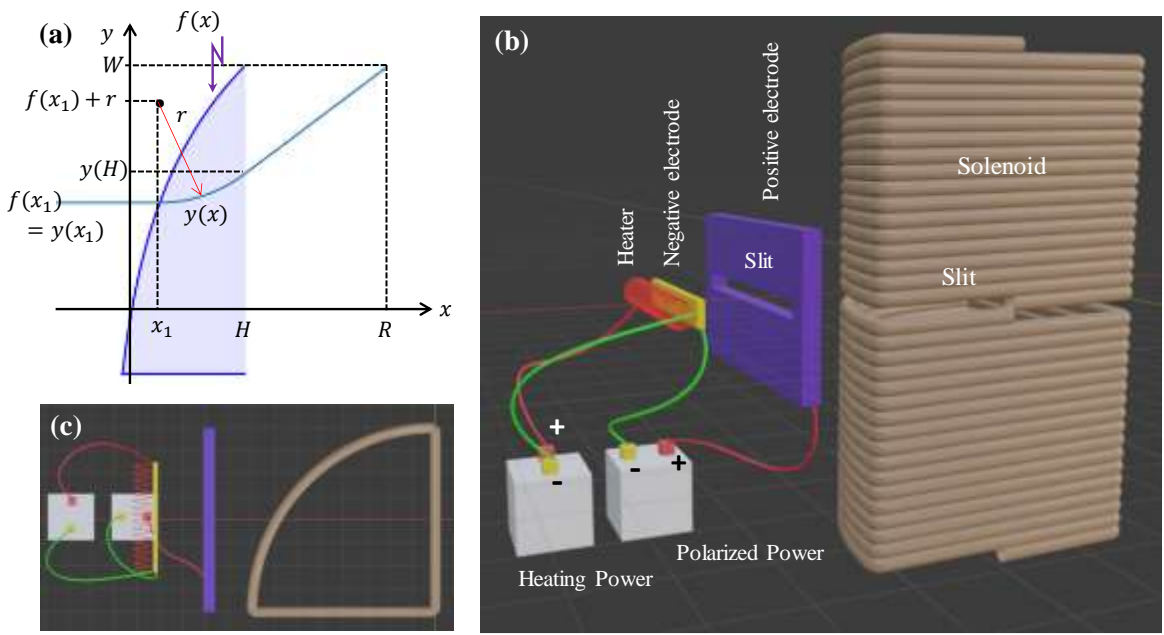

**Figure 1** The magnetic lens concept. (a) Top-side view of the bend in the conductor wire that forms the solenoid (cross section). Three segments make up the solenoid arch: On the left is the $y(x)$ curve, on the bottom is the horizontal straight segment, and on the right is the vertical straight segment. The charged beam comes in from the left and exits from the right. (b) A complete schematic of a magnetic lens as seen from the side. Solenoids have a narrow gap in the center through which the two-dimensional charge beam enters and exits. The gap is so tiny that it has no effect on the magnetic field inside the solenoid. Charge generators, accelerators, and slits for generating two-dimensional loads are located on the beam's inlet side. We assume that the charges are electrons. (c) Top view of the solenoid lens.

We'll start with the most basic geometry of the solenoid lens, as shown in Fig. 1. At the end of this paper, more general geometries will be discussed. To make a magnetic lens, we change the cross section of the solenoid so that the charge beam entering the solenoid is deflected and focused on one point when it exits the solenoid from any point. Various methods for generating charge beams have been discussed by Franken et al [1]. We will use a solenoid with the cross-section shown in Fig. 1(a). The solenoid's surface is divided into three sections: a curve with the equation $y(x)$ on the left, a straight horizontal segment on the bottom, and a straight vertical segment on the right. We will look for the equation for the left-side curve so that all rays coming from the left in a horizontal direction to the right converge at a point to the right of the solenoid.

**Convergent Beam**

The two-dimensional charge beam (forming the $xy$ plane) is traveling at a speed $v$ from the left. Consider the incoming charge at coordinates $(x_1, f(x_1))$. When the charge enters the solenoid cavity, it follows a path in the form of a circular segment with radius $r$ that satisfies

$$r = \frac{mv}{|q|B} \tag{2}$$

Assume the charge is negative and the field is oriented from front to back, causing the track to curve upward as shown in Fig. 1(a). As shown in Appendix 1, the charge will converge at coordinate $(R, W)$ if the curve on the left side of the solenoid satisfies the following equation

$$f(x) = W - r + \sqrt{r^2 - (H-x)^2} - \frac{(H-x)(R-H)}{\sqrt{r^2-(H-x)^2}} \tag{3}$$

The curve's equation is affected by the mass of the charge, the magnitude of the charge, and the magnitude of the magnetic field (contained in $r$). By selecting the appropriate $f(x)$ equation, we can determine the location of the focal point $(W, R)$ from this equation. We can change the focal length by changing $H$ which causes the function $f(x)$ to change slightly but not the form of the function.

The charge path in the solenoid cavity is

$$y(x) = f(x_1) + r - \sqrt{r^2 - (x - x_1)^2} \tag{4}$$

and the charge's path as it exits the solenoid is

$$z(x) = y(H) + a(x - H) \tag{5}$$

where

$$a = -\frac{(H-x_1)}{(y(H)-f(x_1)-r)} \quad (6)$$

Figure 2(a) depicts a parallel beam being bent so that it converges at a focal point behind a magnetic lens. The beam near the top edge is only slightly bent. The beam on the top edge is not deflected because it passes through almost no magnetic field. As shown in Fig. 1(a), the vertical coordinates of the focal point equal the height of the lens cross-section.

If the focal point is very close to the lens's rear side, or $R \approx H$, we get

$$f(x) \approx W - r + \sqrt{r^2 - (H-x)^2} \quad (7)$$

Equation (7) represents the equation of a circle with radius $r$ and a center at coordinate $(H, W - r)$. As a result, the circumference of the solenoid's front segment is a circle. In contrast, if the focal point is very far from the lens, $R \gg H$ and $R \gg \left|W - r + \sqrt{r^2 - (H-x)^2}\right|$, we get the approximation form

$$f(x) \approx -\frac{(H-x)R}{\sqrt{r^2-(H-x)^2}} \quad (8)$$

**Divergent Beam**

Charges of opposite sign deflect downward, as shown in Fig. 2(b). The charge path in the solenoid cavity, as stated in Appendix 2, fulfills the equation

$$y^*(x) = f(x_1) - r + \sqrt{r^2 - (x-x_1)^2} \quad (9)$$

and the charge's path after leaving the solenoid fulfills the equation

$$z^*(x) = y^*(H) + a^*(x - H) \quad (10)$$

where

$$a^* = -\frac{(H-x_1)}{(y^*(H)-f(x_1)+r)} \quad (11)$$

**A point-source beam is transformed into a parallel beam**

We get the following if a point charge is emitted from the focal point and travels in the same direction as the trajectory in Fig. 2(a). Before reaching the solenoid, the charges follow the paths of the parallel negative beam from the left, but in the opposite direction, as shown in Fig. 2(c). When it enters the solenoid, the charge is accelerated according to the equation below

$$\vec{a} = \frac{\vec{F}}{m} = \frac{q(-\vec{v}) \times \vec{B}}{m} \quad (12)$$

Charges of different signs follow the same circular path as the positive charge moving from the left. The charge then leaves the solenoid by traveling horizontally to the left. This means that we can convert the beam leaving the point source into a parallel beam using a solenoid lens.

This is the same behavior as a point light placed at the focal point of a convex lens. The beam propagates in a parallel direction after passing through the lens. If the charges have different signs but the field direction is reversed, the same trajectory is obtained, as shown in Fig. 2(d).

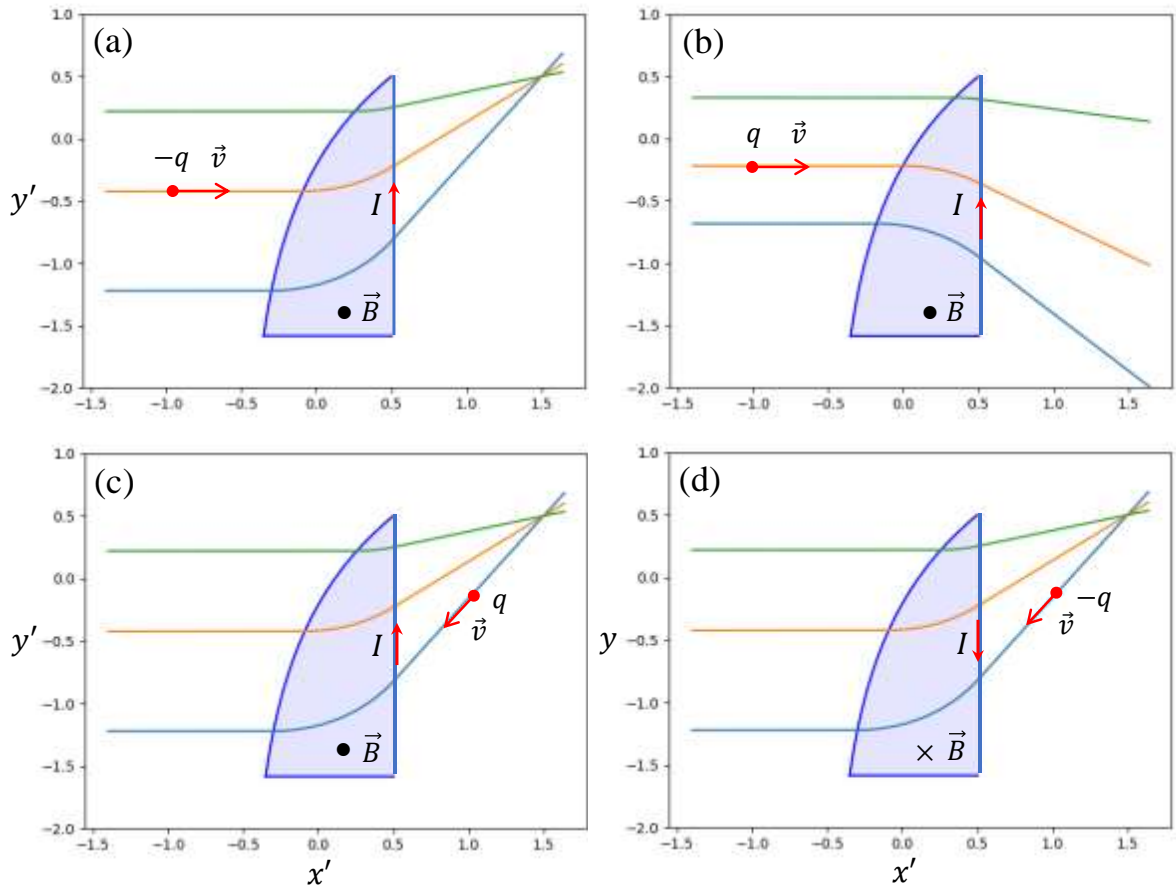

**Figure 2** The charge trajectories in the solenoid lens. (a) and (b) The parallel beam enters the solenoid from the left side, forming a curved trajectory (viewed from above): (a) the negative charge and forward magnetic field will focus the parallel beam, and (b) the positive charge and forward magnetic field will diverge the parallel beams. (c) and (d) are point-source beams that have been converted into parallel beams: (c) positive charge and forward magnetic field and (c) negative charge and backward magnetic field (d). We used $H' = 0.5$, $R' = 1.5$, and $W' = 0.5$ in the simulations.

## Concept of "Mass Spectrometer"

If the radius of the charge path in the solenoid cavity is changed to $r'$, for example, due to changes in the particles' field, charge, or mass, the equation for the circular path inside the solenoid cavity becomes

$$y^{**}(x) = f(x_1) + r' - \sqrt{r'^2 - (x - x_1)^2} \tag{13}$$

The charge exits the solenoid at the new coordinate $(H, y^{**}(H))$ where

$$y^{**}(H) = f(x_1) + r' - \sqrt{r'^2 - (H - x_1)^2} \tag{14}$$

After leaving the solenoid, the charge moves in free space in a straight line with a slope equal to the slope at the point $(H, y^{**}(H))$, i.e.

$$a^{**} = -\frac{(H - x_1)}{(y^{**}(H) - f(x_1) - r')} \tag{15}$$

The straight line equation that the particle follows after leaving the solenoid is

$$z^{**}(x) = y^{**}(H) + a^{**}(x - H) \tag{16}$$

At the horizontal position $R'$, the charge reaches a height $z^{**} = W$, satisfying the equation

$$W = y^{**}(H) + a^{**}(R' - H)$$

or

$$R' = -\frac{(y^{**}(H) - f(x_1) - r')}{(H - x_1)}\left(W - f(x_1) - r' + \sqrt{r'^2 - (H - x_1)^2}\right) + H \tag{17}$$

Figures 3(a) to (c) show the positions of $R'/r$ at various $r'/r$ values for particles originating from different $y_1$. The difference in the value of $r'$ causes the charge to exit the solenoid in a different direction even though it enters from the same point. As the beam enters the solenoid at lower points, its direction becomes increasingly different.

The beams entering the solenoid at different $y$ positions are no longer focused at a single point, as shown in Fig. 3(d)-(f). If we place an ion detector on the back side of the solenoid, we find that in the first case, when the particle path radius is $r$, we will detect a very high intensity at coordinate $(R, W)$. The intensity at the coordinate $(R, W)$ is no longer maximal when the radius of the path in the solenoid becomes $r'$. Lower intensities are obtained by moving the detector about the $(R, W)$ coordinate in the horizontal direction.

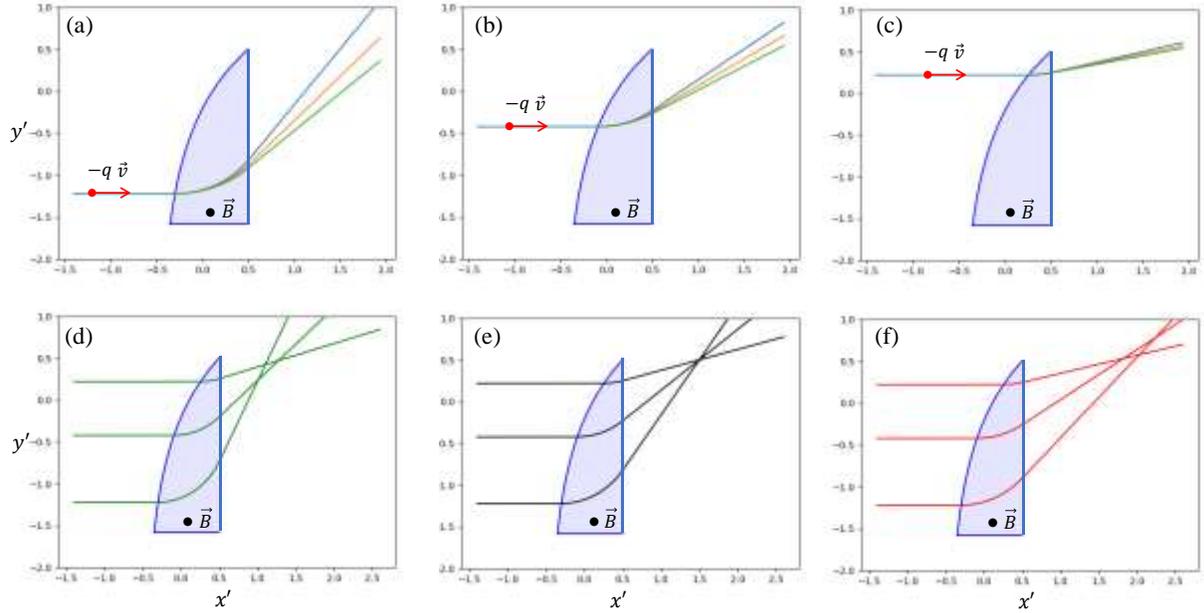

**Figure 3** Illustrations of charge trajectories as the path radius in the cavity is changed. The magnetic field, charge mass, and charge speed can all be adjusted to alter the trajectory radius. In Fig. (a)-(c) we used $r'/r = 1.0$ (top path), $r'/r = 1.1$ (middle path), and $r'/r = 1.2$ (bottom path). Other parameters used include $H' = 0.5$, $R' = 1.5$, and $W' = 0.5$. Figures (d)-(f) show the paths of the incoming beams at various $y$- coordinates demonstrating that the beams are not focused on a single point. In figure (d), we used $r'/r = 0.9$, in figure (e) we used $r'/r = 1.0$, and in figure (f) we used $r'/r = 1.15$.

If the particle beam has the same charge and speed but different masses, the path radius changes according to Eq (2). The focal point is no longer the $(R, W)$ coordinate. However, we can control the magnetic field to be $B'$ in such a way that the radius of the charge path inside the solenoid cavity is exactly the same as that given by Eq. (1), i.e.

$$r = \frac{m'v}{|q|B'} \qquad (18)$$

As a result, the charge beam is again focused on the $(R, W)$ coordinate. This behavior is a mass spectrometer concept.

We can determine the mass of the charge by adjusting the magnetic field in the cavity of the solenoid so that the charge is focused at the coordinate $(R, W)$. Controlling the field is as simple as adjusting the current flowing through the solenoid so that the mass of the particles matches the equation

$$m' = \frac{|q|\mu_0 n I_0 r}{v} \qquad (19)$$

Figure 4 is an illustration of a mass spectrometer using a solenoid lens. The inset is an illustration of the intensity of charge detected by the detector when the current in the solenoid is varied. The peak of the curve represents the condition when the beam is focused at the detector position. Another location with a smaller intensity represents a condition when the charge is no longer focused due to a current (magnetic field) so that the radius of the passage in the solenoid cavity changes.

The charge and speed of the particles entering the solenoid are assumed to be constant. A velocity selector, similar to the front part of a mass spectrometer, can be used to control the particle speed. The mass spectrometer is divided into two sections: the velocity selector, which contains both electric and magnetic fields, and the deflection chamber (the solenoid cavity) which contains only the magnetic fields.

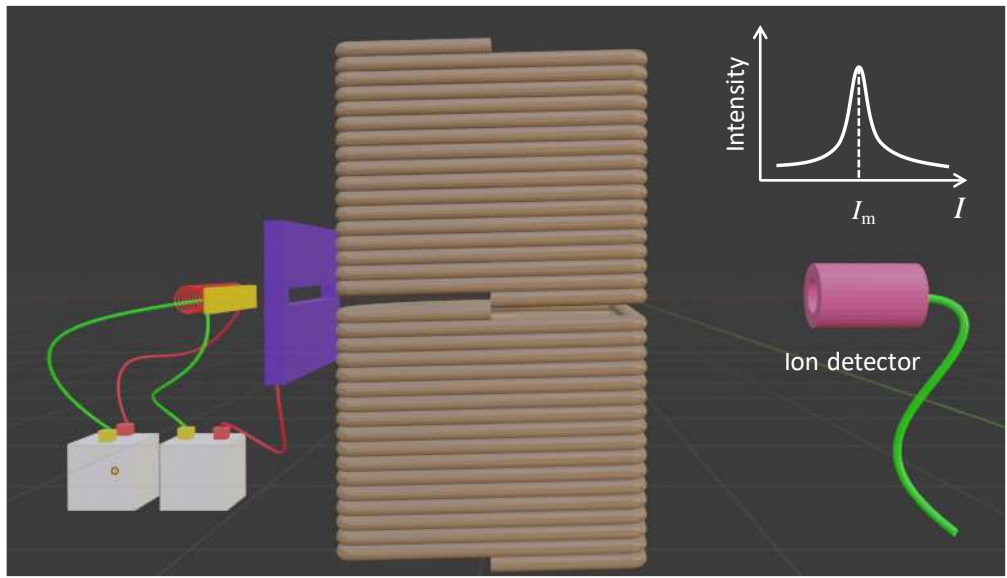

**Figure 4** Mass spectrometer concept. The inset shows an estimate of the detector's intensity as the current flowing through the solenoid is scanned from small to large.

**General Surfaces**

Consider a solenoid lens bounded by arbitrary shaped surfaces. For the sake of simplicity, assume that the beam enters the solenoid from the bottom and exits at the top. The function $f(x)$ defines the lower side of the solenoid circumference, while the function $g(x)$ defines the upper side. At $x = W$, the functions $f(x)$ and $g(x)$ intersect. We make a vertical straight line on the left side of the solenoid surface. Furthermore, we assume that the beam converges at coordinate $(W, R)$. For a more detailed explanation, see Fig. 5.

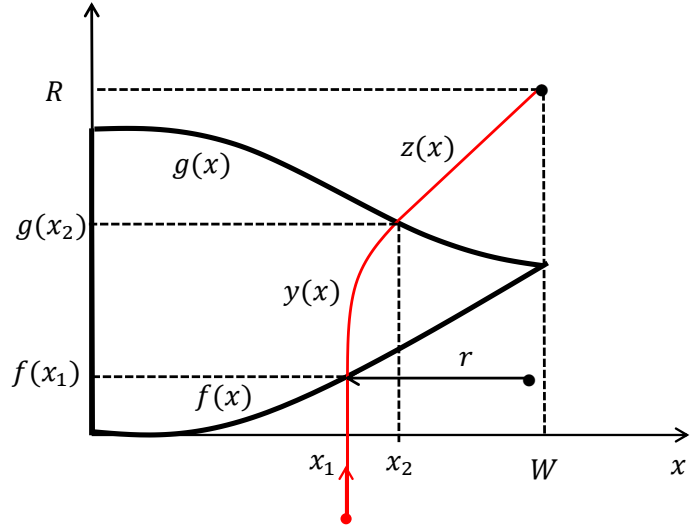

**Figure 5** Solenoid lens with arbitrary cross-sectional shape.

Consider a charge reaching the solenoid's front surface at coordinate $(x_1, f(x_1))$. The charge moves in circular paths within the solenoid cavity, satisfying the equation

$$y(x) = f(x_1) + \sqrt{r^2 - (x - (x_1 + r))^2} \qquad (20)$$

The equation for $g(x)$ that causes all beams to be focused at coordinate $(W, R)$ is

$$g(x_2) = f(x_1) + r\sqrt{1 - s'^2} \qquad (21)$$

where

$$s' = \frac{(W'-x'_1-1) - \sqrt{(R'-f'(x_1))^4 + (R'-f'(x_1))^2(W'-x'_1-1)^2 - (R'-f'(x_1))^2}}{(R'-f'(x_1))^2 + (W'-x'_1-1)^2} \qquad (22)$$

$$x_2 = rs' + (x_1 + r) \qquad (23)$$

and $\gamma' = \gamma/r$ for any variable $\gamma$. The derivation of Eqs. (20)-(23) is given in Appendix 3.

We can deduce from Eq. (21) that for every equation of the solenoid's front side, $f(x)$, there is always an equation for the back side, $g(x)$, so that all the beams collect at one point behind the solenoid.

Let us now look at some of the following special cases..

**Special case 1**: $f(x) = 0$.

We can get easily

$$s' = \frac{(W'-x'_1-1)-\sqrt{R'^4+R'^2(W'-x'_1-1)^2-R'^2}}{R'^2+(W'-x'_1+1)^2} \tag{24a}$$

$$g'(x_2) = \sqrt{1-s'^2} \tag{24b}$$

**Special case 2**: $f(x) = Ax$ where $A$ is a constant.

We can easily show

$$s' = \frac{(W'-x'_1-1)-\sqrt{(R'-Ax'_1)^4+(R'-Ax'_1)^2(W'-x'_1-1)^2-(R'-Ax'_1)^2}}{(R'-Ax'_1)^2+(W'-x'_1-1)^2} \tag{25a}$$

$$g'(x_2) = Ax'_1 + \sqrt{1-s'^2} \tag{25b}$$

**Special case 3**: $f(x) = Ax^2$ where $A$ is a constant.

We can easily show

$$s' = \frac{(W'-x'_1-1)-\sqrt{(R'-A^*x'^2)^4+(R'-A^*x'^2)^2(W'-x'_1-1)^2-(R'-A^*x'^2)^2}}{(R'-A^*x'^2)^2+(W'-x'_1-1)^2} \tag{26a}$$

$$g'(x_2) = A^*x'^2 + \sqrt{1-s'^2} \tag{26b}$$

where $A^* = Ar$.

Figure 6 depicts an example beam propagation for the three preceding cases with $R' = 1.4$, $W' = 1.0$, $A = 0.5$, and $A^* = 0.25$. The magnetic field is oriented backward, and the charge is negative. The function $f(x)$ that represents the front side of the solenoid cross-section appears to have a unique counterpart function $g(x)$ to the back side of the cross-section. This implies that there are an infinite number of cross section shape options for focusing the load on $(R, W)$ coordinates.

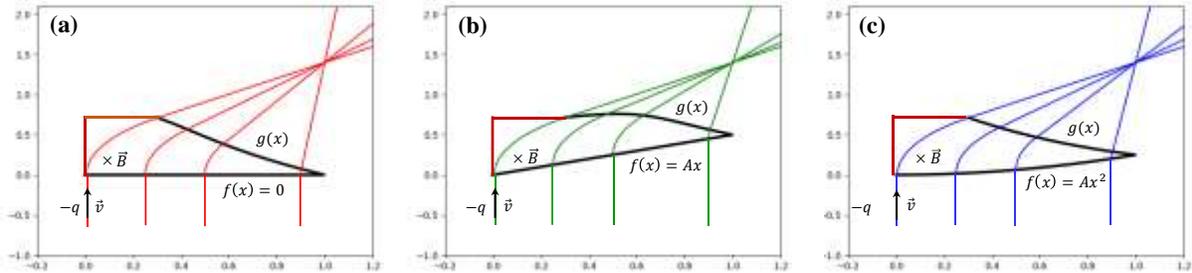

**Figure 6** Example of the path of the beam passing through the lens of the solenoid where the bottom surface (where the parallel charge enters) satisfies the equation: (a) $f(x) = 0$, (b) $f(x) = Ax$ where $A = 0.5$ and (c) $f(x) = Ax^2$ where $Ar = 0.25$. Other parameters that have been used are $R' = 1.4$ and $W' = 1.0$.

In principle, making a solenoid with arbitrary cross-sections is not difficult. The wire is wound on a rod with a cross-section equal to the desired cross-section. The coil of wire with the desired length is then removed from the rod.

**Conclusion**

We can design a magnetic lens to focus the parallel ion beam to the desired point by using an ideal solenoid and taking advantage of the field properties produced by the solenoid. There are an infinite number of curvature solenoid circumference options for focusing the ion beam parallel to a single point. Using this property, we present the concept of a simple mass spectrometer by measuring the intensity of the ion captured by the detector at a specific point.

**Appendix 1**

The circular path when the charge is in the solenoid cavity has a center point at coordinates $(x_1, f(x_1) + r)$ so the path equation is $(x - x_1)^2 + (y(x) - f(x_1) - r)^2 = r^2$ or

$$y(x) = f(x_1) + r \pm \sqrt{r^2 - (x - x_1)^2} \tag{A1.1}$$

where $x_1 \leq x \leq H$. The right side of the solenoid cross section is at coordinate $x = H$. The charge leaves the solenoid at coordinates $(H, y(H))$. Since this point is still on the circular path, we get

$$y(H) = f(x_1) + r \pm \sqrt{r^2 - (H - x_1)^2} \tag{A1.2}$$

Since $y(H) < f(x_1) + r$, so that the charge can exit the solenoid on the right side (not form a full circle in the solenoid cavity), we choose the solution with a negative sign on the last term on the right side of equation (A1.2) so that

$$y(H) = f(x_1) + r - \sqrt{r^2 - (H - x_1)^2} \tag{A1.3}$$

After leaving the solenoid, the particle moves in a straight line in free space with a slope equal to the slope at point $(H, y(H))$. The slope on a circular path at any coordinate $(x, y)$ satisfies the equation

$$\frac{dy}{dx} = -\frac{(x - x_1)}{(y - f(x_1) - r)} \tag{A1.4}$$

Thus, the slope at the point $(H, y(H))$ which then becomes the slope of the particle path after leaving the solenoid is

$$a = -\frac{(H - x_1)}{(y(H) - f(x_1) - r)} \tag{A1.5}$$

The equation of the straight line that the particle takes after leaving the siolenoid is

$$z(x) = y(H) + a(x - H) \tag{A1.6}$$

We want particles entering from any point on the front side of the solenoid to converge at coordinate $(R, z(H) = W)$. We can say that the coordinate $(R, W)$ is the focal point. The resulting equation is

$$W = y(H) + a(R - H) \tag{A1.7}$$

Substituting $y(H)$ from equation (A1.3) and a from equation (A1.5) into equation (A1.7) we get

$$W = y(H) - \frac{(H-x_1)}{(y(H)-f(x_1)-r)}(R - H)$$

to yield

$$f(x_1) = W - r + \sqrt{r^2 - (H - x_1)^2} - \frac{(H-x_1)(R-H)}{\sqrt{r^2-(H-x_1)^2}} \tag{A1.8}$$

For any $x_1$ on the left side of the solenoid cross section, we obtain the general equation

$$f(x) = W - r + \sqrt{r^2 - (H - x)^2} - \frac{(H-x)(R-H)}{\sqrt{r^2-(H-x)^2}} \tag{A1.9}$$

## Appendix 2

For charges of opposite sign, the charge deflects downward as shown in Figure 2(b). The center of the circular path is at coordinates $(x_1, f(x_1) - r)$. The equation for the circle of charge paths inside the solenoid cavity is

$$(x - x_1)^2 + (y^* - f(x_1) + r)^2 = r^2 \tag{A2.1}$$

The charge leaves the solenoid at the coordinate $(H, y'_2)$ that satisfies the equation

$$(H - x_1)^2 + (y^*(H) - f(x_1) + r)^2 = r^2$$

or

$$y^*(H) = f(x_1) - r \pm \sqrt{r^2 - (H - x_1)^2} \tag{A2.2}$$

But $y^*(H) \geq f(x_1) - r$ so in equation (A2.2) we choose the positive sign on the last term so

$$y^*(H) = f(x_1) - r + \sqrt{r^2 - (H - x_1)^2} \tag{A2.3}$$

As it leaves the solenoid cavity, the gradient of the charge path is

$$a^* = -\frac{(H-x_1)}{(y^*(H)-f(x_1)+r)} \tag{A2.4}$$

This gradient becomes the gradient of the charge's straight path as it leaves the solenoid. Thus, the equation for the position of the charge becomes

$$z^*(x) = y^*(H) + a^*(x - H) \tag{A2.5}$$

## Appendix 3

Consider the charge hitting the surface that the solenoid is at coordinates $(x_1, f(x_1))$. Within the solenoid cavity, the particles move in circular paths that satisfy the equation

$$y(x) = f(x_1) + \sqrt{r^2 - (x - (x_1 + r))^2} \qquad (A3.1)$$

This beam intersects the function $g(x)$ at coordinates $(x_2, g(x_2))$. Since these coordinates are also on the circular path (the end of the circular path when it leaves the solenoid), we obtain

$$g(x_2) = y(x_2) = f(x_1) + \sqrt{r^2 - (x_2 - (x_1 + r))^2} \qquad (A3.2)$$

The slope of the circle curve in equation (A3.1) at any point is

$$a = -\frac{(x - (x_1 + r))}{(y(x) - f(x_1))} \qquad (A3.3)$$

The slope at the point $(x_2, g(x_2))$ where the charge leaves the solenoid is

$$a = -\frac{(x_2 - (x_1 + r))}{(y(x_2) - f(x_1))}$$

$$= -\frac{(x_2 - (x_1 + r))}{(g(x_2) - f(x_1))} \qquad (A3.4)$$

The equation of the straight line leaving the point $(x_2, g(x_2))$ which is the path of the charge as it leaves the solenoid is

$$z(x) = g(x_2) + a(x - x_2) \qquad (A3.5)$$

The directions of all the beams leaving the solenoid pass through the coordinates $(W, R)$ we obtain the equation

$$R = g(x_2) + a(W - x_2)$$

$$= g(x_2) - \frac{(x_2 - (x_1 + r))}{(g(x_2) - f(x_1))}(W - x_2) \qquad (A3.6)$$

Substituting equation (A3.2) into equation (A3.6) we get

$$R = f(x_1) + \sqrt{r^2 - (x_2 - (x_1 + r))^2} - \frac{(x_2 - (x_1 + r))}{\sqrt{r^2 - (x_2 - (x_1 + r))^2}}(W - x_2) \qquad (A3.7)$$

For simplicity, let's define

$$s = x_2 - (x_1 + r) \qquad (A3.8)$$

We can prove that $s < 0$. Substituting equation (A3.8) into equation (A3.7) we get

$$(R - f(x_1))\sqrt{r^2 - s^2} = r^2 - s(W - (x_1 + r)) \qquad (A3.9)$$

Let's normalize all parameters in equation (A3.9) with respect to $r$ so that we can find solutions more easily. The normalized equation is

$$(R' - f'(x_1))\sqrt{1 - s'^2} = 1 - s'(W' - x'_1 - 1) \tag{A3.10}$$

where $\gamma' = \gamma/r$ for any variable $\gamma$. The solution for $s'$ is [Wolframalpha]

$$s' = \frac{(W' - x'_1 - 1) - \sqrt{(R' - f'(x_1))^4 + (R' - f'(x_1))^2 (W' - x'_1 - 1)^2 - (R' - f'(x_1))^2}}{(R' - f'(x_1))^2 + (W' - x'_1 - 1)^2} \tag{A3.11}$$

Having obtained $s'$, we can calculate $x_2$ using equation (A3.8) and we can calculate $g(x_2)$ using equation (A3.2). We can prove easily that

$$g(x_2) = f(x_1) + r\sqrt{1 - s'^2} \tag{A3.12}$$

## Acknowledgements

Not applicable.

## Funding
No funding has been used for the production of this research.

## Declarations of Competing interests
The authors declares that they have no competing interests

**Mikrajuddin Abdullah**
http://orcid.org/ 0000-0001-5098-4712